\def\hcorrection#1{\advance\hoffset by #1 }
\def\vcorrection#1{\advance\voffset by #1 }

\documentstyle[11pt]{article}

\vcorrection{-0.70in}
\hcorrection{-0.40in}
\begin{document}

\vspace{0.3in}

\begin{center}
{\Large
\bf Simulation of Cavity Flow \\ by the Lattice
      Boltzmann Method }
\end{center}

\vspace{0.3in}

\begin{center}
Shuling Hou$^{1,3}$, Qisu Zou$^{2}$,
Shiyi Chen$^1$, Gary D. Doolen$^1$, Allen C. Cogley$^3$
\end{center}

\vspace{0.3in}

${}^{1}${\footnotesize  Center for Nonlinear Studies and Theoretical
Division, Los Alamos National Laboratory,} \\
\hspace*{0.3in}{\footnotesize Los Alamos, NM 87545}

${}^{2}${\footnotesize  Department of
Mathematics, Kansas State University, Manhattan, KS 66506.}

${}^{3}${\footnotesize  Department of Mechanical Engineering,
Kansas State University, Manhattan, KS 66506.}

\vspace{0.3in}

\begin{center}
{\bf Abstract}
\end{center}

A detailed analysis is presented to demonstrate the capabilities of
the lattice Boltzmann method. Thorough comparisons with other numerical
solutions for the two-dimensional, driven cavity flow show that the lattice
Boltzmann method gives accurate results over a wide range of Reynolds numbers.
Studies of errors and convergence rates are carried
out. Compressibility effects are quantified for different maximum velocities,
and parameter ranges are found for stable simulations. The paper's
objective is to stimulate further work using this relatively new approach for
applied engineering problems in transport phenomena utilizing parallel
computers.

\section{Introduction}

  Lattice gas automata (LGA) and its later derivative, the lattice Boltzmann
equation method (LBE), are relatively new approaches that utilize parallel
computers to study transport phenomena.
Since the first two-dimensional model representing incompressible
Navier-Stokes equations was proposed by Frisch, Hasslacher, and Pomeau (FHP)
in 1986 [1], LGA have attracted much attention as promising
methods for solving a variety of partial differential equations and modeling
physical phenomena [2, 3, 4, 5].

A lattice gas is constructed as a simplified, fictitious microworld
in which space, time and velocities are all discrete. In general, a lattice gas
consists of a regular lattice with particles residing on the nodes.
A set of Boolean variables ${n_i}({\bf x},t) ( i=1,\cdots ,b$) for describing
the particle occupation is defined, where $b$ is the number of
directions of particle velocity at each node. Starting from an initial state,
the configuration of particles at each time step evolves in two sequential
steps: (a) streaming, where each particle moves to the nearest node in
the direction of its velocity; and (b) colliding, which occurs when particles
arriving at a node interact and possibly change their velocity directions
according to scattering rules. For simplicity, the exclusion principle
(no more than one particle is allowed at a given time and node moving in
a given direction) is imposed for memory efficiency and leads to the
Fermi-Dirac equilibrium distribution. The strategy of the lattice gas is
two-fold: a) to construct a model as simple as possible of the microworld to
permit simulations of a system composed of many particles and b) to capture the
essential features of real collision processes between particles such that,
for long times and large scales, macroscopic transport phenomena are captured.

 That the evolution of particles on an artificial lattice can simulate the
macroscopic behavior of fluid flow is based on the following facts: the
macro-dynamics of a fluid is the result of the collective behavior of many
particles in the system and details of the microscopic interactions are not
essential. Changes in molecular interactions affect transport properties such
as viscosity, but do not alter the basic form of the macroscopic equations as
long as the basic conservation laws and necessary symmetries are satisfied
[2, 3].

Due to the microscopic nature and local interaction between particles, the
lattice gas approach possesses some unique advantages. The scheme is absolutely
stable; boundary conditions are easy to implement; the model is ideal for
massively parallel computing and the code is simple. The lattice
gas method also contains some problems such as non-Galilean invariance, due
to the existence of a density-dependent coefficient in the
convection term of the Navier-Stokes equation, an unphysical
velocity-dependent pressure and inherent statistical noise that requires a
spatial (or time) averaging to obtain smooth macroscopic quantities. To avoid
some of these inherent problems, several lattice Boltzmann (equation) models
have been proposed [6, 7, 8, 9, 10, 11]. The main feature of the LBE is to
replace the particle occupation variables, $n_i$, (Boolean variables) by the
single-particle distribution functions (real variables) $f_i=\langle n_i
\rangle $, where $ \langle \;  \rangle $ denotes a local ensemble average,
in the evolution equation, i.e. the lattice Boltzmann equation.

 The lattice Boltzmann
equation as a numerical scheme was first proposed by McNamara and Zanetti [6].
In their model, the collision operator is the same as in the LGA. Higuera,
Jimenez and Succi [7, 8] introduce a linearized collision operator that is a
matrix and has no correspondence to the detailed collision rules. Statistical
noise is completely eliminated in both models; however, the other
problems remain since the equilibrium distribution is still Fermi-Dirac. The
lattice Boltzmann model proposed by Chen {\it et al}. [9, 11] and Qian
{\it et al}. [10] abandons Fermi-Dirac statistics and provides the freedom
required for the equilibrium distribution to satisfy isotropy, Galilean
invariance and to possess a velocity-independent pressure. Their models apply
the single relaxation time approximation first introduced by Bhatnager, Gross
and Krook in 1954 [12] to greatly simplify the collision operator. This model
is called the lattice Boltzmann BGK model.

Compared with the lattice gas approach, the lattice Boltzmann method is more
computationally efficient using current parallel computers. Applications
have been done using both methods on hydrodynamics [13, 14, 15, 16], flow
through porous media [17, 18], magnetohydrodynamics [19, 20], multiphase flow
[21, 22, 23, 24] and the reaction-diffusion equation [25, 26, 27]. Collected
papers and applications of lattice gas and lattice Boltzmann methods can be
found in [4, 5, 28, 29].

Despite these studies on various problems, thorough quantitative
investigations of the method have not been published. In the present work, the
lattice Boltzmann BGK model (LBGK) is used to solve for the viscous flow in a
square, two-dimensional cavity driven by shear from a moving wall for Reynolds
numbers up to 10,000. Detailed comparisons between the LBGK and traditional
methods are presented. The compressibility error and the convergence
rate of the method are discussed. The objective of this paper is to analyze the
accuracy and physical fidelity of the lattice Boltzmann BGK method and to
stimulate further studies using the lattice Boltzmann approach.

Section 2 presents a technical synopsis of the lattice Boltzmann model used in
this paper that will enhance the general reader's understanding of this
simulation method. More technical details are given in the Appendix for those
who want to use the lattice Boltzmann method. The lattice Boltzmann simulation
of driven cavity flow is discussed in Section 3 and thoroughly compared with
results from other numerical methods. Section 4 studies the numerical
errors in lattice Boltzmann simulations due to lattice size and
compressibility. Section 5 is devoted to comparisons between the square
lattice and the triangular (FHP) lattice. The limit of relaxation time for
these two models is explored. The final section contains concluding remarks.

\section{Two-Dimensional Square Lattice Boltzmann Model}

In this section an outline is given of the procedures of the lattice
Boltzmann simulation. A square lattice with unit spacing is used on which each
node has 8 nearest neighbors connected by 8 links (see Figure 1A in
the Appendix). Particles can only reside on the nodes and move to their
nearest neighbors along these links in the unit time step. Hence, there are
two types of moving particles. Particles of type 1 move along the axes
with speed $|{\bf e_1}|=1$ and particles of type 2 move along the
diagonal directions with speed $|{\bf e_2}|=\sqrt{2}$. Rest particles with
speed zero are also allowed at each node. The occupations of the three types of
particles is represented by the single-particle distribution function,
$f_{\sigma i}({\bf x},t) $, where subscripts $\sigma$ and $i$ indicate the type
of particle and the velocity direction, respectively. The distribution
function, $f_{\sigma i}({\bf x},t)$,
is the probability of finding a particle at node ${\bf x}$ and time
$t$ with velocity ${\bf e}_{\sigma i}$. The particle distribution function
satisfies the following lattice Boltzmann equation:
\begin{equation}
f_{\sigma i}({\bf x}+{\bf e}_{\sigma i},t+1)-f_{\sigma i}({\bf x},t)
= \Omega_{\sigma i},
\label{eq:1}
\end{equation}
where $\Omega_{\sigma i}$ is the collision operator representing the rate of
change of the particle distribution due to collisions.
According to Bhatnagar, Gross and Krook (BGK) [12], the collision operator is
simplified by the single time relaxation approximation. Hence, the lattice
Boltzmann BGK equation is
\begin{equation}
 f_{\sigma i}({\bf x}+{\bf e}_{\sigma i},t+1 )
-f_{\sigma i}({\bf x},t)
=-\frac{1}{\tau} [f_{\sigma i}({\bf x},t)-f_{\sigma i}^{(0)}({\bf x},t)],
\label{eq:2}
\end{equation}
where $f_{\sigma i}^{(0)}({\bf x},t) $ is the equilibrium distribution at
${\bf x}, t$ and $\tau $ is the single relaxation time which controls the rate
of approach to equilibrium.
The density per node, $ \rho $, and the macroscopic velocity, ${\bf u}$,
are defined in terms of the particle distribution function by
\begin{equation}
\sum_{\sigma} \sum_{i} f_{\sigma i}=\rho,
\label{eq:3}
\end{equation}
and
\begin{equation}
 \sum_{\sigma} \sum_{i} f_{\sigma i}{\bf e}_{\sigma i} =\rho {\bf u}.
\label{eq:4}
\end{equation}
The equilibrium distribution can be chosen in the following form for particles
of each type:
\begin{eqnarray}
f_{0 1}^{(0)}=\frac{4}{9}\rho[1-\frac{3}{2}u^2], \mbox{\hspace{1.65 in}}
\nonumber\\
f_{1 i}^{(0)}=\frac{1}{9}\rho[1+3({\bf e}_{1 i}\cdot
{\bf u})+\frac{9}{2}({\bf e}_{1 i}\cdot {\bf u})^2-\frac{3}{2}u^2],\nonumber \\
f_{2 i}^{(0)}=\frac{1}{36}\rho[1+3({\bf e}_{2 i}\cdot
{\bf u})+\frac{9}{2}({\bf e}_{2 i}\cdot {\bf u})^2-\frac{3}{2}u^2].
\label{eq:5}
\end{eqnarray}
The relaxation time is related to the viscosity by
\begin{equation}
 \nu=\frac{2 \tau -1}{6} ,
\label{eq:6}
\end{equation}
where $\nu$ is the kinematic viscosity.
The detailed derivation of the LBGK model is given in the Appendix.

Having chosen the appropriate lattice size and the characteristic velocity
for the LBE system, the viscosity, $\nu$, of the problem can be calculated for
a given Re number and then the relaxation time is determined by the formula
above. Starting from an initial state of $ f_{\sigma i}({\bf x}, t)$, the
density and velocity fields and hence the equilibrium distribution function
can be obtained. In each time step, the updating of the particle
distribution can be split into two substeps: collision and streaming. It is
irrelevant which one is the first for a long time run. The collision process
at position ${\bf x}$ occurs according to the right hand side of the
Boltzmann equation given as Eq.~(\ref{eq:2}). The resulting particle
distribution at
${\bf x}$, which is the sum of the original distribution and the collision
term, is then streamed to the nearest neighbor of ${\bf x},{\bf x}+{\bf e}_
{\sigma i}$, according to the particle velocity ${\bf e}_{\sigma i}$.
The updating procedure is terminated for steady state problems when certain
criteria are reached. The method can be used for transient problems, but this
will not be discussed in this paper.

The boundary condition commonly used at the solid wall of a fluid simulation is
the no-slip condition for which the velocities vanish at the wall. This is
implemented in the lattice gas and lattice Boltzmann methods with the
bounce-back rule in which all particles hitting the wall are reflected back in
the direction from which they came.

Another lattice model commonly used in two-dimensional lattice gas and lattice
Boltzmann
simulations is the triangular lattice (FHP model) [1, 2]. This is a two-speed
(0 and 1) model in which the lattice constant (link) is equal to one.
Simulations of cavity flow are also performed in this paper using this model.
Comparisons between FHP and square lattice are discussed in Section 5. Two
commonly used models for three-dimensional simulations are the 24-velocity
FCHC [2] and the 14-velocity cubic models [10, 14].

\section{Cavity Simulation}

  The problem considered is two-dimensional viscous flow in a cavity
governed by the Navier-Stokes equations. An incompressible fluid is
bounded in a square enclosure and the flow is driven by the uniform
translation of the top boundary. The fluid motion generated in a cavity is an
example of closed streamline problems that are of theoretical importance
because they are part of broader field of steady, separated flows.
The literature is abundant for this flow configuration that shows rich vortex
phenomena at many scales depending on the Reynolds number, Re.
Numerical methods for solving the Navier-Stokes equations are often
tested and evaluated on cavity flows because of the complexity of the flows.

  Most numerical solutions of two-dimensional
cavity flow [32-40] use a vorticity-stream function formulation and discretize
the incompressible, steady linear or nonlinear Navier-Stokes
equations by finite difference [32-35], multigrid [36, 38, 39] and finite
element [40] methods and their variations [37]. Earlier work was reviewed by
O. Burggraf [32] where his numerical solutions of the nonlinear Navier-Stokes
equations for Reynolds number up to 400 showed a large primary vortex and two
secondary vortices in the lower corners. The later studies of A. S. Benjamin
and V. E. Denny [35], U. Ghia, K. N. Ghia and C. T. Shin [36], R. Schreiber
and H. B. Keller [37] show that
tertiary vortices are formed near the bottom corners for higher Reynolds
numbers. The present results using the lattice Boltzmann method are compared
with those done by Vanka [38], Schreiber and Keller [37], Ghia {\it et al}.
[36] and Zhou {\it et al}. [39]. Ghia {\it et al}. obtained numerical solutions
up to Re=10,000 with a 257$ \times $257 grid using the coupled strongly
implicit multigrid method and vorticity-stream function formulation. Their work
is the most comprehensive study of cavity flow to date.

  The present simulation uses Cartesian coordinates with the origin located
at lower left corner. The top wall is moving from left to right with velocity
$U$. The cavity has 256 lattice units on each side. Initially the velocities
at all nodes, except the top, are set to zero. The x-velocity of the top is
$U$ and the y-velocity is zero. Uniform initial particle density is imposed
such that the moving particle 1 has a density fraction of $d=\frac{\rho}{9}
=0.3$ per direction. The moving particle 2 has a density fraction of
$\frac{d}{4}$ per direction, and the rest particle has a density of $4d$.
Therefore, the total density per node is $\rho =2.7$. Using the uniform
density distribution and velocities given above, the equilibrium particle
distribution function, $f_i$, is calculated according to Eq.~(\ref{eq:5}). The
evolution of $f_i$ can then be found by a succession of streaming and
collision-like processes. After streaming, the velocity of the top lid is
reset to the uniform initial velocity. At the end of each streaming and
collision process cycle, the particle distribution function, $f_i$, at the top
is set to the equilibrium state and the bounce-back boundary conditions are
used on the three stationary walls. The two upper corners are singular points
which are considered as part of the moving lid in the simulations, but tests
shown there is little difference if these two points are treated as fixed wall
points. The uniform velocity of the top wall used in the simulations is
$U$=0.1. The compressibility effects are discussed in Section 4.4. The
Reynolds number used in the cavity simulation is defined as $Re=$U$ L/\nu$,
where $U$ is the uniform velocity of the top plate, $L$ is the edge length of
the cavity and $\nu$ is the kinematic viscosity that is related to the single
relaxation time as given in Eq.~(\ref{eq:6}). All the results are
normalized to allow comparisons between the present work and other
results based on a unit square cavity with unit velocity of the top boundary.

  Steady-state solutions for cavity flow are obtained using the lattice
Boltzmann method for Re=10, 100, 200, 400, 1,000, 2,000, 5,000, and 7,500.
The Re=10,000 case is also run on a 256$ \times $256 lattice, but steady state
cannot be reached because bifurcation takes place somewhere between Re=7,500
and 10,000. The results for Re=10,000 oscillate between a series of different
configurations. For this reason the results presented in this paper are those
for Re up to 7,500. The dependent variables of stream function, velocity,
pressure and vorticity are calculated using the particle distribution
function, $f_i$. The dependent parameter of the drag coefficient of the driving
wall is discussed also.

\subsection{Stream function}

  Figure 1 (a-g) shows plots of the stream function for the Reynolds numbers
considered. It is apparent that the flow structure is in good agreement with
the previous work of Benjamin and Denny [35], Schreiber and Keller [37] and
Ghia {\it et al}. [36]. These plots give a clear picture of the overall flow
pattern and the effect of
Reynolds number on the structure of the steady recirculating eddies in the
cavity. In addition to the primary, center vortex, a pair of counterrotating
eddies of much smaller strength develop in the lower corners of the cavity at
higher values of Re. At Re=2000, a third secondary vortex is seen in the upper
left corner (it is generated at a critical Re of about 1,200 according to
[35] and that agrees with the results of the present work). For Re $\ge $5,000,
a tertiary vortex in the lower right hand corner appears.
A series of eddies with exponentially decreasing strength in the lower corners
has been predicted [36]. Due to the compressibility effect of the LBE
(discussed in Section 4.4), the tertiary vortex shown in the Figure 1 (g)
oscillates.

For low Re (e.g. Re=10), the center of the primary vortex is located
at the midwidth and at about one third of the cavity depth from the top (see
Figure 1.a). As Re increases (Re=100), the primary vortex center moves towards
the right and  becomes increasing circular. Finally, this center moves
down towards the geometric center of the cavity as the Re increases and becomes
fixed in its x location for Re $\ge$ 5,000. The movement of
the vortex center location versus Re is shown in Figure 2 along with the
results given by Ghia {\it et al}. [36].

  To quantify these results, the maximum stream function value for the primary
vortex and the minimum values for the secondary vortices along with the
x and y coordinates of the center of these vortices are listed in Table 1. All
the results presented use a uniform top velocity $U=0.1$ except for Re=100
where U=0.01 is used. The reason is discussed in Section 4.4. Also
listed are results selected from previous work [36-39]. Previous results agree
with each other for Re $ \le$ 1000, but vary for higher values of Re.
The results of the present work and that of Ghia {\it et al}.
[36] for stream function values agree within 0.2 \% for all values
of Re (Re=2,000 data was not given by [36]).
The locations of the vortex centers predicted by the lattice Boltzmann method
also agree well with those given by Ghia {\it et al}. [36].

Unlike finite-difference or finite-element methods that start from the
steady-state, partial differential equations, the present method is a direct
modeling method that evolves into steady state. The time to reach
steady-state depends on the lattice size, the values of Re and the driving
velocity, $U$. For all the cases run in this paper, steady-state is reached
when
the difference between the maximum values of the stream function for successive
10,000 steps (process cycles) is less than $10^{-5}$.
Considering the kinetic, direct, compressible and unsteady nature of the
lattice Boltzmann method, the excellent agreement with entirely different
methods such as Ghia {\it et al}. [36] is quite encouraging.

The minimum values of stream function and the center of the secondary vortex
in the upper left corner for Re=5,000 and 7,500 are listed in Table 2. These
results also show good agreement with Ghia {\it et al}. [36].

\subsection{Velocity profiles}

  Velocity components along a vertical and horizontal center line for several
values of Re are shown in Figure 3. The velocity profiles change from curved
at lower values of Re to linear for higher Re values. The near linear profiles
of the velocity in the central core of the cavity indicate the uniform
vorticity region generated in the cavity at higher values of Re. These results
agree with those from previous studies [34, 36, 37].

\subsection{Vorticity}

  The plots of vorticity in Figure 4 (a-g) show that the steady cavity flow
within closed streamlines at high Re consists of a central, inviscid core of
nearly constant vorticity with viscous effects confined to thin shear
layers near the walls. Batchelor [30] predicted these results from his
model for separated eddies in a steady flow.
As the Re increases, several regions of high vorticity gradients (indicated by
concentration and wiggle of the vorticity contours) appear within the cavity.
The thinning of the wall boundary layers with increasing Re is evident from
these plots, although the rate of this thinning is very slow for Re $\ge $
5,000. The values of vorticity at the center of the primary vortex for
different Re are listed in Table 3. These values closely agree with the results
of Ghia {\it et al}. [36] and approach the analytical value of 1.886 for an
infinite Reynolds number calculated by Burggraf [32] using Batchelor's model.

\subsection{Pressure}

  Figure 5 (a-g) displays the pressure deviation contours for the present
simulations. Since only the pressure gradient appears in the Navier-Stokes
equation, the values of pressure can deffer linearly.
These plots are in good agreement with the static pressure given by Burggraf
[32] (note that the top wall in [32] moves from right to left which is
opposite to that in the present simulation). The pressure in [32] is obtained
by integrating the Navier-Stokes equation given the velocity field, while the
pressure in the lattice Boltzmann method satisfies the equation of state of
the isothermal gas where it is proportional to the density. The observed
agreement between these very different approaches demonstrates that the lattice
Boltzmann BGK model is valid for simulating incompressible flow.
By examing the closed contours in the pressure plots, it is seen that the
inviscid core grows with increasing values of Re. In the opposite limit
of Re approaching zero, pressure becomes a harmonic function and the contours
cannot be closed but must end on the boundaries [32]. The present results
support this last requirement.

\subsection{Drag on the top}

The drag force and the drag coefficient of the moving wall is calculated here
for Re values considered in the study. The stress on the moving wall is given
by the Newton's formula as
\[\tau_{yx}=\mu \frac{\partial u}{\partial y}, \]
where $u$ is the x component of velocity and $\mu$ is the kinetic viscosity.
The drag force on this surface, $F_d$, is defined as
\[ F_d=\int_0^L \tau_{yx} dx=,
\int_0^L \mu \frac{\partial u}{\partial y}dx=\mu \sum_{i=2}^{i=n_x-1}
\frac{u(i,2)-u(i,1)}{\triangle y } \triangle x, \]
where $n_x$ is the grid number in the x direction, $L$ is the length of the
square cavity and $\triangle y=\triangle x =\frac{L}{(n_x-1)}$ are the spacing
of the lattice. The drag coefficient is then written as
\[C_d= \frac{F_d}{\bar{\rho} U^2 L},\]
where $\bar{\rho}$ is the average density and $U$ is the velocity of the top.
The drag coefficient decreases as Re increases, as found in other laminar
flow configurations. This can be seen by introducing the dimensionless
quantities
\[u'=\frac{u}{U}, \;\;\; x'=\frac{x}{L},\;\;\; y'=\frac{y}{L}. \]
The drag coefficient can then be expressed as
\[ C_d=\frac{U \int_0^1 \mu \frac{\partial u'}{\partial y'} dx'}
{\bar{\rho} U^2 L}=\int_0^1 \frac{\mu}{\bar{\rho} U L}
\frac{\partial u'}{\partial y'} dx'
=\frac{1}{Re} \int_0^1 \frac{\partial u'}{\partial y'} dx'. \]

 The results of drag and drag coefficient for different values of Re are
listed in Table 4 and the latter is plotted in Figure 6. The plot shows that
the relation between drag coefficients and Re satisfies the above formula.
There is no data available on drag and drag coefficient from other methods for
comparison.

\section{Error Analysis}

\subsection{Sources of errors}

 There is no analytic solution for cavity flow. Results from the work described
in this paper are compared with the numerical solutions obtained by several
other methods. Differences are found between the results of previous work,
especially for higher values of Re. Several of these authors state that the
data for the secondary vortices are less reliable due to corner singularities
and/or roundoff errors [37], mesh-size limitations [34] or because the values
of the stream function in the corners are small and, in some cases, below the
convergence accuracy of the calculations [38].

  Before the final results listed in Table 1 were obtained, the results from
lattice Boltzmann simulations were very close to the results given by Ghia
{\it et al}. [36] for Re$\ge$ 1,000 only. The properties of the secondary
vortices were less satisfactory for Re less than 1,000. The secondary vortex
of the lower left corner for Re=100, whose stream function is a small quantity
of the order of $10^{-6}$, was not detected by the lattice Boltzmann
method used in the initial simulations. Also, the secondary vortex in the lower
right corner for the same Re, whose stream function is of the order of
$10^{-5}$, did not match corresponding results of other investigators. Although
these are not major features, it was important to investigate the cause of
these discrepancies.

  The theoretical assumptions of the present method are the Boltzmann transport
equation plus the single relaxation time approximation of the collision term.
As long as the macroscopic properties of fluid vary slowly enough in space and
time compared with microscopic particle dynamics, collisions should maintain
approximately the local equilibrium such that the assumptions of molecular
chaos by Boltzmann and single relaxation time by BGK are valid for problems of
fluid dynamics. The possible reasons for the ``errors'' in the present
simulations may be categorized as follows:

1. The simulations done for different values of Re on a 256$\times $ 256
lattice used single precision arithmetic. It is possible that roundoff error
could be accumulated.

2. The small compressibility effect presented in the LBE simulations
may cause differences when compared with models where compressibility is zero.

3. The lattice size used here may still be too coarse to resolve all the small
scale phenomena.

4. The time step at which the simulation is terminated may not be large enough
to represent steady state.

5. The integration methods used in the calculation of the stream function may
introduce errors.

The errors caused by 1. and 4. can be avoided by using double precision
floating-point arithmetic and running for longer times. These experiments did
not change the result for weak vortices on the smaller scales. What follows
are subsections investigating the rest of the error sources.

\subsection{Effect of lattice size}

  To test the effect of lattice size, simulations for Re=1,000 are done
on the following lattice configurations: 33 $\times $ 33,
65 $\times $ 65, 129 $\times $ 129, and 513 $\times $ 513.
The driving velocity used is kept at  $U=0.1$.
Two relative velocity errors were calculated according to the following
formula:

\[ E1=\frac{\sum_{x,y} |u_1-u_0|+|v_1-v_0|}
              {\sum_{x,y}|u_0|+|v_0|}, \]

\[ E2=\frac{\sqrt{\sum_{x,y} (u_1-u_0)^2 + (v_1-v_0)^2}}
              {\sqrt{\sum_{x,y} (u_0)^2 + (v_0)^2}}, \]
where $u $, $v$ are the $x$ and $y$ components of the velocity, respectively.
The subscript 0, 1 indicate the $513 \times 513$ and the coarser grain
lattice, respectively. Velocities on different grids are taken at corresponding
positions, while the sums are taken over the entire lattice. E1 and E2 are
two common relative velocity errors in the $L1$ and $L2$ norm sense,
respectively, based on the highest resolution lattice (513).

The results for $E1$ and $E2$ are plotted logarithmically in Fig. 7 and
listed in Table 5.
The quantities $E1$ and $E2$ are calculated also according to the formula
above where the 0 and 1 indicate two successive lattice sizes. The results are
similar to that shown in Fig. 7. It is clear from Fig. 7 that the convergence
rate is approximately first order in space. This result is different from
other works [15, 41] where a second-order convergence rate is claimed. In [15]
the problem is a decaying Taylor vortex flow with periodic boundary conditions
(no solid wall) and in [41] exact boundary conditions for the particle
distribution function are used instead of the bounce-back condition implemented
at the wall in the present work. The first order convergence rate observed here
may be due to the bounce-back condition used on the stationary walls.
This observation is confirmed by a recent paper [42] in which the bounce-back
boundary condition is shown to be a first-order approximation for a no-slip
wall. It is shown also in [42] that higher order accuracy can be achieved
by improving the implementation of boundary conditions.

Better resolution is obtained as the number of lattice nodes increases.
However, computer time grows with lattice number because more nodes are updated
and the time to reach steady-state is much longer. The time steps required
to reach steady state for different lattice sizes are listed in Table 6.
Beyond these facts, a simulation run on a $513 \times 513 $ lattice for Re=100
did not improve the method's ability to predict the secondary vortex in the
lower, left corner.

\subsection{Integration error}

  The most important features of the cavity flow are related to the stream
function. The stream function used by Ghia {\it et al}. [36] (and others) was
the primary variable of the solution. In the lattice Boltzmann model, however,
the primary variable is the particle distribution function, $f_i$. The
velocity at each site is calculated from $f_i$ and the stream function is
obtained by integrating the velocities.

  To investigate the error caused by integration, three integration rules
(rectangular, trapezoidal and Simpson) are used for Re=100. The results from
all three rules are of the same order of accuracy if the integrations are
taken in the same direction (of course the trapezoidal and Simpson rules are
higher-order integration schemes than the rectangular rule and hence produce
less error). Significantly
different results are obtained by integrating from the four different
directions, namely integrate $u$ along $y$ from top to bottom, integrate $u$
along $y$ from bottom to top, integrate $v$ along $x$
from left to right and integrate $v$ along $x$ from right to left.
Theoretically, they should all give the same value for the stream function.

{}From a numerical point of view, the integration should be
taken from the smaller scale, otherwise the smaller scale would be dawn into
roundoff error. However, in this particular case, integration from bottom
to top contains significant error. The reason can be seen from the error
formula.  If the trapezoidal rules is used:
\[I(f)=\int_a^b f(x) dx \approx h[\frac{1}{2} f(a)+f(a+h)+\cdots+\frac{1}{2}
f(b)]=I_n(f), \]
where $h=\frac{b-a}{n}$ and $n$ is the number of subdivisions within
$[a,b]$, the integration error can be expressed as
\[E_n(f)=I(f)-I_n(f)=-\frac{h^2(b-a)}{12}f''(c_n). \]
Here $c_n$ is a point between $a$ and $b$. The above formula can be
improved by the asymptotic error formula [43]:
\[E_n(f)\approx -\frac{h^2}{12}[f'(b)-f'(a)].\]
The factor, $\frac{h^2}{12}$, in the present case is about $1.3 \times
10^{-6}$. The error then depends on the derivatives at the end points. In the
case of integration taken from top to bottom or from bottom to top, the two
derivatives have opposite signs and the error is enhanced. In addition, the
value of the derivative on the top is large. Numerical tests show that
integration from top to bottom gives inaccurate values ($1.96e^{-3}$ and
$1.66e^{-3}$ for the minimum stream function on the lower left and right
corners, respectively). The integration from the bottom to the top predicts
another vortex of order$ 10^{-3}$ on the upper left corner that does not exist
for Re=100 and is caused by errors (due to both integration and
compressibility). On the
other hand, integration from left to right or from the right to left gives the
same signs on the two end derivatives, therefore decreasing the integration
error. Since the left corner vortex is smaller than that on the right, the
integration taken from left to right gives better results ($-1.76e^{-6}$on
left and $-1.10e^{-5}$ on right) than the integration taken from the right to
the left ($-2.4e^{-5}$ on left and $-1.13e^{-5}$ on right). The conclusion is
that using a trapezoidal rule and integrating the velocity component, $v$,
from left to right gives the most accurate integral. The error is then of
order of $10^{-9}$. The Simpson rule has about the same accuracy.

\subsection{Compressibility effect}

It has been shown that the present LBE model represents the Navier-Stokes
equation in the incompressible limit (see Appendix). But in the LBE
simulation, the density cannot be a constant (otherwise pressure change cannot
be described). It is important to find the effect of compressibility on the
present solution.

One quantity that represents compressibility is the mean variation of density.
The mean density is defined as
\[\bar{\rho}=\frac{\sum_i \rho(x_i, t)}{N}, \]
where $N$ is the total number of nodes. The mean variation of density is
given by
\[\bigtriangleup= \frac{1}{\bar{\rho}} \sqrt{\frac{\sum ((\rho -\bar{\rho})^2)}
{N}}. \]
For Re=100, this mean density fluctuation, $\bigtriangleup$, is calculated for
$U=0.1, U=0.05, U=0.01$ and listed in Table 7 along with the Mach number,
$M=\frac{u}{c_s}$, where $c_s=\frac{1}{\sqrt{3}}$ is the speed of sound for
the present model. The table shows that
\[ \bigtriangleup (U=0.05) \approx \frac{1}{4} \bigtriangleup (U=0.1), \]
and
\[ \bigtriangleup (U=0.01) \approx \frac{1}{25} \bigtriangleup (U=0.05). \]
These results agree with the known relationship [16] that $\bigtriangleup$
is proportional to $M^2$. (This relation can be seen from the dimensionless
incompressible Navier-Stokes equations.)

 The compressibility effect can also be examined for the cavity flow problem
as follows. In the steady case, the continuity equation represented
by LBE is
 \[ \nabla \cdot (\rho {\bf u} )=0, \]
due to a non-constant $\rho$. The velocity $\bf u $ does not satisfy the
incompressible continuity condition given by
\[ \nabla \cdot {\bf u }=0. \]
It is from this equation that the stream function can be defined using
$u=\frac{\partial \psi}{\partial y}$ and $v=-\frac{\partial \psi}{\partial x}$,
where $\psi$ is the stream function. There is actually no exact
definition for the stream function in LBE.
Given a discrete velocity field obtained from the LBE calculation,
an approximation of the stream function for the incompressible
flow with $ \nabla \cdot ({\bf u })=0 $ needs to be constructed.
The stream function definition written as
 \[\psi =\int -vdx +udy \]
is still used to calculate the stream function.
When integrating in only the x- or y-direction, the integral becomes
\[\psi =\int_0^y udy, \]
or
\[\psi =- \int_0^x vdx. \]
In the case of incompressible flow in a cavity, the boundaries coincide with
the zero stream function. The integrals then take the form
\[\psi =\int_0^L udy = \int_0^L vdx =0, \]
where $L$ is the total length of the wall. For an incompressible model,
integration of $u$ (or $v$) along the $y$ (or $x$) direction from one edge of
the cavity gives a theoretical value of the stream function at another edge
of zero. In the actual computations, the stream function at the wall will
not exactly equal zero because of roundoff and integration error.
Due to the additional effects of compressibility in the LBE method,
if the stream function is calculated by integrating $v$ from the left to the
right edge of the cavity, the values of the stream function on
the right wall would indicate the error caused by
compressibility, roundoff and integration errors. Since the trapezoidal rule
gives the same results for integrations taken from opposite directions if there
is no roundoff error, the roundoff error is found by comparing the
values of the stream function on the left and right wall taken from opposite
directions (the other sources of error, compressibility and integration, are
the same for these two integrals).
This error is less than about $10^{-9}$. The integration error
is of the order of $10^{-9}$ as discussed above. Therefore, the maximum
and the mean value of $\psi$ at the right wall can be computed as an indicator
of error due to compressibility if this value is larger than $10^{-8}$. The
mean and maximum stream function at the right edge of the cavity is defined,
respectively, as
\[S_a=\sqrt{\frac{\sum(\psi^2(n_x,j))}{n_y}},\]
and
\[ S_m = \max_j | \psi(n_x,j)| ,\]
where $n_x=n_y=256$ is the number of nodes in the x- and y-direction,
respectively.
These values are calculated for $U=0.1, U=0.05$ and $U=0.01$ for Re=100 and
listed in Table 8. Again, the results show that $S_a$ and $S_m$ are
proportional to $M^2$. The compressibility error calculated by this argument
can be used as a quantitative measure of the compressibility of the LBE method.
The change of compressibility error with Re is calculated for $U=0.1$ and
listed in Table 9. The error caused by compressibility does not vary
much with Re number. Actually with increased Re the error is
slightly decreased, but is still of the same order of magnitude.

It is clear that the error caused by compressibility has about the same order
effect as the small scale phenomena in the cavity flow for low values of Re.
By choosing
the direction of integration for the stream function carefully, predictions
of the small vortices are obtained. Furthermore, the compressibility error can
be reduced by using smaller velocities at the top wall. The results for Re=100
in Table 1 are calculated using $U=0.01$, while other Reynolds number use
$U=0.1$. However, the time steps required to reach steady-state for smaller
top velocity increases dramatically as seen in Table 10. To overcome the
compressibility error in the present LBE model, a new incompressible LBE model
for steady-state flow has been developed and will be published in another
paper [44].

\section{Triangular Lattice (FHP) Versus Square Lattice}

Simulations for cavity flow are also carried out on a triangular lattice (FHP).
There are two types of particles on each node of a FHP model: rest particles
and
moving particles with unit velocity ${\bf e}_i$ along 6 directions.
In analogy to the procedures used for the square lattice in the Appendix, the
equilibrium distributions for the FHP model are given as
\[ f_{0}^{(0)}=d_0-\rho u^2 = \alpha \rho -\rho u^2 , \]
\[ f_{i}^{(0)}=d+\frac{1}{3}\rho[({\bf e}_{i}\cdot {\bf u})+2({\bf e}_{ i}
\cdot {\bf u})^2-\frac{1}{2}u^2] \\
=\frac{\rho-\alpha \rho}{6} +\frac{1}{3}\rho[({\bf e}_{i}\cdot {\bf u})+
2({\bf e}_{i} \cdot {\bf u})^2-\frac{1}{2}u^2]. \]
If the ratio of rest and moving particle is defined as
\[\lambda =\frac{d_0}{d}, \]
the pressure is determined by the following
isothermal equation of state:
\[ p=3 d=\frac{(1-\alpha) \rho}{2}= \frac{3}{\lambda +6} \rho, \]
and the speed of sound is
\[c_s^2= \frac{1-\alpha}{2}=\frac{3}{\lambda +6}. \]
The viscosity is related to the relaxation time through an equation of the
form
\[\nu=\frac{2 \tau -1}{8}. \]
Theoretically, the relaxation time, $\tau$, cannot be lower than 0.5 for a
positive viscosity. To reach higher Re, the relaxation time can be
lowed. Tests on a 128 $\times $ 128 lattice with a maximum velocity of $U=0.1$
show that a critical value for $\tau$ exists. Above this value, the simulation
is smooth and reasonable physical patterns for the cavity flow are
seen in the real-time plots. However, below this critical value, some
nonphysical patterns appear.  Further reduction in the value of $ \tau$ would
cause the simulation to be terminated by numerical blow up. Define the critical
value of
$\tau$ as the lowest limit for the relaxation time that gives physically
correct results. This limit varies with the ratio, $\lambda $, maximum
velocity,
$U$, and the problem studied. If $\lambda $ is increased, the speed of the
sound will be decreased and the Mach number is then increased if the velocity
is unchanged. Table 11 lists the lowest relaxation time, $\tau_{min} $, and
hence the highest Re number, $Re_{max}$, obtained for different $\lambda $,
along with their speed of sound, $c_s$, and Mach number, $M$. Table 11 shows
that the highest Re can be increased by increasing $\lambda$. However, the
compressibility error is also increased.

Table 12 lists results for the same conditions as in Table 11, but for slightly
different initial boundary conditions on the density. In case 1 of Table 11,
the initial density on each node of the wall is the same uniform
distribution as that for interior nodes. In case 2 of Table 12,
the initial density on the wall is set equal to zero. It is clear
that when $ \lambda =1 $, the lowest $\tau$ is much higher than that in Table
11, but for large $ \lambda$ the differences of $\tau_{min}$ between these two
cases are diminished. The rest particles in the LBE method play the role of a
particle reservoir. When the macroscopic velocity is higher, rest particles can
be turned into moving particle and {\it vice versa}. Higher values of
$\lambda=d_0/d$ mean a larger fraction of rest particles in the density behave
like a fluid that is less rigid and more flexible (the compressibility is
higher). When $\lambda=1 $ in case 2, the lowest value of $\tau $ for stable
results is 0.5668. However, setting $\tau=0.5667 $ would make the computation
blow up immediately due to the large initial density gradient on the wall and
the relatively small fraction of rest particles. The lowest limit of $\tau$
does not depend on the lattice size. Changing the maximum velocity, $U$, does
change the lowest $\tau $ slightly. However, the highest Re numbers obtainable
by this approach are much lower than that for $U=0.1$ (see Table 13 for case
2).

The square lattice corresponds to $\lambda=\frac{d_0}{d_1}=4$. Tests on a 128
$\times $ 128 square lattice with the maximum velocity of $U=0.1 $ show that
the value
of $\tau $ cannot be smaller than 0.507 (Re=5485) for cavity flow. Hence a
simulation run on a 256 $\times $ 256 lattice with $U=0.1$ can reach Re=10,000
($\tau=0.50768 $) which is about the highest limit of Re on this size of
lattice for cavity flow. Using small $U$ did not produce a further reduction of
$\tau$. The square lattice is better than the FHP lattice in the cavity
flow simulations because the former can reach higher values of Re than the
latter for the same maximum velocity and lattice size. Since the
boundaries of the cavity are fitted better using the square lattice than FHP
lattice, the formation of the vortices is more gentle in the simulation
process using a square lattice than a FHP lattice. The ranges of parameters
presented in this section are consistent with the results of linear stability
analysis of LBE method without boundaries [45].

\section{Conclusions}

The lattice Boltzmann method is a derivative of the lattice gas automata method
and therefore inherits from the LGA some of its advantages over
traditional computational methods. It is parallel in nature due to the
locality of the transport of particle information, so it is well
suited to massively parallel computing. Due to the kinetic description of the
lattice Boltzmann method, it is easy to handle the complex boundary
conditions and properties of a fluid system, such as flow through porous media
and multi-phase flow. One important improvement due to the LBE method is that
it can fully recover the Navier-Stokes equations at the macroscopic level
including Galilean invariance and a velocity-independent pressure. However,
there is a trade-off. The lattice Boltzmann method no longer has the pure
Boolean operation and numerical stability guaranteed by LGA.

Detailed study of the cavity flow problem using the lattice Boltzmann
method has shown that the method is accurate compared with conventional methods
using the same mesh size. This verification produces confidence to apply the
method to other complex systems. All aspects of the present work
such as boundary conditions, parameter ranges, lattice size and compressibility
effects are important when the method is applied to other problems. The
following remarks are in order:

1. The proper implementation of the boundary conditions is crucial for the
lattice Boltzmann simulation. Various boundary conditions such as
periodic, particle bounce-back, wind tunnel and constant flux conditions are
commonly used for different situations in LBE. It is important that the
boundary conditions applied for the simulation represent the correct physical
problem. In the cavity simulation, for example, besides
the uniform top velocity and no-slip conditions on the wall, the mass
must be conserved globally. Any violation of this restriction will produce
nonphysical results. Be aware that some improper boundary conditions can give a
qualitatively reasonable flow but lead to quantitatively incorrect results.

2. The range of parameters for the model is explored for the cavity
simulations. Parameters such as the lattice size, maximum velocity,
the ratio of rest and moving particle and the single relaxation time
are adjustable in LBE. The lattice size should be chosen so that a good
resolution for all scales in the problem can be obtained at an affordable cost.
The maximum velocity used in a simulation should be
properly small for a low Mach number and hence low compressibility
requirement and for the validation of the equilibrium distribution which is an
expansion of small velocity. For the Chapman-Enskog expansion to be valid,
the spatial gradients of density and velocity should be small also. Since the
maximum velocity and lattice size are limited, the single relaxation time needs
to be small to achieve the higher Reynolds numbers. It is found that the lowest
relaxation time leading to stable simulations depends on the ratio of rest and
moving particles, the maximum velocity and the problem. To obtain a reliable
simulation, the relaxation time should be chosen not too close to the lowest
limit for the problem under investigation. On the other hand, since the lattice
spacing in the LBE is unit, the parameter, $\tau $, can be understood as a mean
free path. Therefore, $\tau $ should be small enough compared with macroscopic
characteristic length scale. This is a necessary condition that the microscopic
statistics of the LBE will approach the Navier-Stokes equations as shown in
the multiscale expansion (see the Appendix).

3. The compressibility effect may become important when physical quantities of
the smallest scale in an incompressible flow is comparable to the
compressibility error. Using a smaller maximum velocity can reduce this error.
However, it is probably not practical to predict small scales on the order of
$10^{-8}$ or smaller by the present LBE method.

4. The square lattice is better than the triangular lattice
(FHP) in two-dimensional simulations because the former can reach
higher values of Re number for the same lattice size and maximum velocity.

5. The computer time used in the simulations is not compared carefully with
other methods since the LBE includes transient effects in this problem and
hence is not economical compared with the multigrid method. There is no doubt,
however, that the method can simulate unsteady and other complex problems on a
parallel computer with time comparable, if not superior, to other
methods.

Lattice gas and lattice Boltzmann methods are relatively new approaches for
transport phenomena. It is apparent that further research on both theoretical
and practical aspects is needed. Implementation of higher order boundary
conditions, models for better resolving the small scale phenomena,
applications in new fields and to discover new physics, improvement of
thermodynamical models and careful studies for three-dimensional geometries
are challenges for future research.

{\bf Acknowledgments}

Discussions with Drs. Li-shi Luo, Daryl Grunau, Xiaowen Shan, Jim Sterling,
Yue Hong Qian and Norman Zabusky are appreciated. This work is supported by
the Department of Energy at Los Alamos National Laboratory. The authors wish to
acknowledge the Advanced Computing Laboratory of Los Almos National Laboratory
for providing us use of the connection machines, CM-200
and CM-5, located at this facility. Hou and Cogley from Kansas State University
gratefully appreciate support from IBM, the National Science Foundation
(Grant DDM-9113780) and the Kansas Space Grant Consortium.

\newpage

{\bf Reference}

\begin{description}

\item[1.] U. Frisch, B. Hasslacher and Y. Pomeau, {\em Phys. Rev. Lett.} {\bf
56},
1505 (1986).

\item[2.] U. Frisch, D. d'Humi\`{e}res, B. Hasslacher P. Lallemand, Y. Pomeau
and J-P Rivet, {\em Complex Systems } {\bf 1}, 649 (1987).

\item[3.] S. Wolfram, {\em J. Stat. Phys.} {\bf 45}, 471 (1986).

\item[4.] {\em Lattice Gas Methods for Partial Differential Equations,} edited
by G.D. Doolen (Addison-Wesley Publishing Company, 1989).

\item[5.] {\em Lattice Gas Methods: Theory, Applications and Hardware,} edited
by G.D. Doolen (Physica D, {\bf 47}, 1991).

\item[6.] G. McNamara and G. Zanetti, {\em Phys. Rev. Lett.} {\bf 61}, 2332
(1988).

\item[7.] F. Higuera and J. Jimenez, {\em Europhys. Lett.} {\bf 9}, 663 (1989).

\item[8.] F. Higuera and S. Succi, {\em Europhys. Lett.} {\bf 8}, 517 (1989).

\item[9.] S. Chen, H. Chen, D. Martinez and W.H. Mattaeus, {\em Phys.
Rev. Lett.} {\bf 67}, 3776 (1991).

\item[10.] Y. Qian, D. d'Humi\`{e}res and P. Lallemand, {\em Europhys. Lett.}
{\bf17} (6), 479 (1992).

\item[11.] H. Chen, S. Chen and W.H. Matthaeus, {\em Phys. Rev. A}
{\bf 45}, 5339 (1992).

\item[12.] P.L. Bhatnagar, E.P. Gross and M. Krook, {\em Phys. Rev.}
{\bf 94}, 511 (1954).

\item[13.] S. Succi, R. Benzi and F. Higuera, {\em Physica D} {\bf 47}, 219,
(1991).

\item[14.] S. Chen, Z. Wang, X. Shan and G.D. Doolen, {\em J. Stat. Phys.}
{\bf 68}, 379 (1992).

\item[15.] G. McNamara and B. Alder, ``Lattice Boltzmann simulation of high
Reynolds number fluid flow in two dimensions'' {\em Microscopic Simulations of
Complex Hydrodynamic Phenomena }, edited by M. Mareschal and B.L. Holien,
(Plenum, New York, 1992).

\item[16.] D.O. Mattinez W.H. Matthaeus, S. Chen and D.C. Montgomery,
``Comparison of spectral Method and lattice Boltzmann simulations of
two-dimensional hydrodynamics'' {\em Phys. Fluids A }, (in press).

\item[17.] D. Rothman, {\em J. Geophys. Res.} {\bf 95}, 8663 (1990).

\item[18.] S. Chen, K. Diemer, G.D. Doolen, K. Eggert, C. Fu, S. Gutman
and B.J. Travis,  {\em Physica D} {\bf 47}, 72, (1991).

\item[19.] H. Chen and W.H. Matthaeus, {\em Phys. Rev. Lett.} {\bf 58}, 1845
(1987).

\item[20.] D. Martinez, S Chen and W.H. Matthaeus, "Lattice Boltzmann
Magnetohydrodynamics", submitted to {\em Physics of Plasma}, (1993).

\item[21.] D. Rothman and J. M. Keller, {\em J. Stat. Phys.} {\bf 52},
1119 (1988).

\item[22.] J.A. Somers and P.. Rem, {\em Physica D} {\bf 47}, 39 (1991).

\item[23.] D. Grunau, S. Chen and K. Eggert {\em Phys. Fluids A } {\bf 5} (10),
2557 (1993).

\item[24.] X. Shan and H. Chen, {\em Phys. Rev. E} {\bf 47} (3), 1815 (1993).

\item[25.] D. Dab, A. Lawniczak, J.-P. Boon and R. Kapral,
{\em Phys. Rev. Lett.} {\bf 64}, 2462 (1990).

\item[26.] R. Kapral, A. Lawniczak and P. Masiar, {\em Phys. Rev. Lett.}
{\bf 66}, 2539 (1991).

\item[27.] S. Ponce Dawson, S. Chen, and G. Doolen, {\em J. Chem. Phys.}
{\bf 98}, 1514 (1993).

\item[28.] {\em Discrete Kinetic Theory, Lattice Gas Dynamics and Foundations
of Hydrodynamics }, edited by R. Monaco (World Scientific, 1988).

\item[29.] {\em J. Stat. Phys.}, {\bf 68} No. 3/4, 1992, edited by J.P. Boon.

\item[30.] G.K. Batchelor, {\em  Introduction to fluid dynamics,} 147 (1956).

\item[31.] G.K. Batchelor, {\em  J. Fluid Mech.} {\bf 1}, 177 (1956).

\item[32.] O. Burggraf, {\em  J. Fluid Mech.} {\bf 24}, 113 (1966).

\item[33.] F. Pan and A. Acrivos, {\em J. Fluid Mech.} {\bf 28}, 643 (1967).

\item[34.] J.D. Bozeman and C. Dalton, {\em J. Comput. Phys.} {\bf 12}, 348
(1973).

\item[35.] A.S. Benjamin and V.E. Denny, {\em  J. Comput. Phys.} {\bf 33},
340 (1979).

\item[36.] U. Ghia, K.N. Ghia, and C.Y. Shin, {\em  J. Comput. Phys.}
{\bf 48}, 387 (1982).

\item[37.] R. Schreiber and H.B. Keller, {\em J. Comput. Phys.} {\bf 49},
310 (1983).

\item[38.] S.P. Vanka, {\em J. Comput. Phys.} {\bf 65}, 138 (1986).

\item[39.] M. Zhou, C.L. Huang and Q. Zou, ``A Multigrid Method for the
High-Re Solutions of Navier-Stokes Equations in Primitive Variables'' preprint,
(1992).

\item[40.] P. Demaret and M.O. Deville, {\em J. Comput. Phys.} {\bf 95},
359 (1991).

\item[41.] P.A. Skordos, ``Initial and boundary conditions for the lattice
Boltzmann methods'' preprint (1993).

\item[42.] D.P. Ziegler, {\em J. Stat. Phys.} Vol.{\bf 71}, Nos. 5/6, (1993).

\item[43.] K. Atkinson, {\em Elementary Numerical Analysis}, 2nd Ed.,
John Wiley \& Sons, Inc., New York, 173-175 (1993).

\item[44.] Q. Zou, S. Hou, S. Chen and G.D. Doolen, in preparation.

\item[45.] J.D. Sterling and S. Chen, `` Stability analysis of lattice
Boltzmann method'', submitted to {\em  J. Comput. Phys.}, (1993).

\end{description}

{\bf Figure captions}

Figure 1. (a) Stream function for Re=10. Top velocity U=0.1
The center of the primary vortex is at (0.5216, 0.7686).
The center of lower left vortex is at (0.0392, 0.0431).
The center of lower right vortex is at (0.9647, 0.0392).

Figure 1. (b) Stream function for Re=100. Top velocity U=0.01.
The center of the primary vortex is at (0.6196, 0.7373).
The center of lower left vortex is at (0.0392, 0.0353).
The center of lower right vortex is at (0.9495, 0.0353).

Figure 1. (c) Stream function for Re=400. Top velocity U=0.1.
The center of the primary vortex is at (0.5608, 0.6078).
The center of lower left vortex is at (0.0549, 0.0510).
The center of lower right vortex is at (0.8902, 0.1255).

Figure 1. (d) Stream function for Re=1000. Top velocity U=0.1.
The center of the primary vortex is at (0.5333, 0.5647).
The center of lower left vortex is at (0.0902, 0.0784).
The center of lower right vortex is at (0.8667, 0.1137).

Figure 1. (e) Stream function for Re=2000. Top velocity U=0.1.
The center of the primary vortex is at (0.5255, 0.5490).
The center of lower left vortex is at (0.0902, 0.1059).
The center of lower right vortex is at (0.8471, 0.0980).

Figure 1. (f) Stream function for Re=5000. Top velocity U=0.1.
The center of the primary vortex is at (0.5176, 0.5373).
The center of lower left vortex is at (0.0784, 0.1373).
The center of lower right vortex is at (0.8078, 0.0745).

Figure 1. (g) Stream function for Re=7500. Top velocity U=0.1.
The center of the primary vortex is at (0.5176, 0.5333).
The center of lower left vortex is at (0.0706, 0.1529).
The center of lower right vortex is at (0.7922, 0.0667).

Figure 2. The locations of the center of the primary vortex for different
values of Re numbers. The origin is the geometric center of the cavity.

Figure 3. (a) Velocity profiles for u through the geometric center of the
cavity.

Figure 3. (b) Velocity profiles for v through the geometric center of the
cavity.

Figure 4. Vorticity contours of the cavity flow.
           (a) Re=10.
           (b) Re=100.
           (c) Re=400.
           (d) Re=1000.
           (e) Re=2000.
           (f) Re=5000.
           (g) Re=7500.

Figure 5. Pressure deviation contours of the cavity flow.
           (a) Re=10.
           (b) Re=100.
           (c) Re=400.
           (d) Re=1000.
           (e) Re=2000.
           (f) Re=5000.
           (g) Re=7500.

Figure 6. Drag coefficient of top wall versus. Re.

Figure 7. Convergence rate of the LBE method for cavity flow at Re=1000
with top velocity U=0.1.
The errors are calculated relative to results obtained on a $513 \times 513 $
lattice.

\newpage

{\bf Appendix}

This appendix details the derivation that shows how the
Navier-Stokes equations are recovered from a lattice Boltzmann equation on a
square lattice by using the Chapman-Enskog expansion procedure of
kinetic theory. In addition, the equilibrium distribution functions are
obtained to guarantee that the requirements of isotropy, Galilean-invariance
and velocity-independent pressure are satisfied.

On each node of a square lattice there are three types of particle,
namely, a rest particle, a particle moving along perpendicular directions
and a moving particle along diagonal directions. (see Figure 1A).

\vspace{3.5 in}

\begin{center}
Figure 1A  Schematic of a square lattice
\end{center}

The velocity vectors ${\bf e}_{1,i}, {\bf e}_{2,i}$ are defined as
\[ {\bf e}_{1,i}=(cos \frac{i-1}{2} \pi, sin \frac{i-1}{2} \pi) \;\;\;
i=1, \cdots, 4, \]
\[ {\bf e}_{2,i}=\sqrt{2} (cos (\frac{i-1}{2} \pi+\frac{\pi}{4}),
 sin (\frac{i-1}{2} \pi+\frac{\pi}{4})) \;\;\; i=1, \cdots, 4. \]
The symmetric properties of the tensor
\[\sum_i (e_{\sigma i \alpha} e_{\sigma i \beta} \cdots ) \]
are needed in the derivation and given as follows: \\
The odd order of tensors are equal to zero, i.e.
\setcounter{equation}{0}
\begin{equation}
\sum_i {\bf e}_{\sigma i}=0, \;\;\; \sigma =1,2,
\label{eq:l1}
\end{equation}
\begin{equation}
\sum_i e_{\sigma i \alpha} e_{\sigma i \beta} e_{\sigma i \gamma}=0,
 \;\;\; \alpha,\beta,\gamma=1,2,
\label{eq:l2}
\end{equation}
and
\begin{equation}
\sum_i e_{\sigma i \alpha} e_{\sigma i \beta} e_{\sigma i \gamma}
             e_{\sigma i \theta}e_{\sigma i \zeta} =0,\;\;\;
             \alpha,\beta,\gamma,\theta,\zeta=1,2.
\label{eq:l3}
\end{equation}
The second order of tensor satisfies
\begin{equation}
\sum_i e_{\sigma i \alpha} e_{\sigma i \beta}=2 e_{\sigma}^2
\delta_{\alpha \beta},\;\;\; \alpha,\beta=1,2,
\label{eq:l4}
\end{equation}
where
 \[ \delta_{\alpha, \beta}=\left \{\begin{array}{ll}
                              1,\;\;\;\;& \alpha=\beta, \\

                              0,\;\;\;\;& \alpha \neq \beta,
                                  \end{array}
                             \right.\]
and
 \[ e_{\sigma}=\left \{\begin{array}{ll}
                              1,\;\;\;\;& \sigma=1, \\
                              \sqrt{2},\;\;\;\;& \sigma=2,
                                  \end{array}
                             \right.\]
is the length of ${\bf e}_{\sigma i}$. \\
Finally, the fourth order tensor has an expression as
\begin{eqnarray}
\sum_i e_{\sigma i \alpha} e_{\sigma i \beta} e_{\sigma i \gamma}
       e_{\sigma i \theta} =\left \{\begin{array}{ll}
       2 \delta_{\alpha \beta \gamma \theta},\;\;\;\;& \sigma=1,\\
       4 \Delta_{\alpha \beta \gamma \theta}-
       8 \delta_{\alpha \beta \gamma \theta},\;\;\;\;& \sigma=2,
                                    \end{array}
                            \right.
\label{eq:l5}
\end{eqnarray}
where
    \[\delta_{\alpha \beta \gamma \theta}=\left \{\begin{array}{ll}
                              1,\;\;\;\;& \alpha=\beta=\gamma=\theta, \\
                              0,\;\;\;\;& \mbox{otherwise},
                                  \end{array}
                             \right.\]
and
\[ \Delta_{\alpha \beta \gamma \theta}=(\delta_{\alpha \beta}
                                        \delta_{\gamma \theta}
+\delta_{\alpha \gamma} \delta_{\beta \theta}
+\delta_{\alpha \theta} \delta_{\beta \gamma}). \]
Here the subscripts, $\sigma$, denote the types of particle; $i$ indicates the
directions of particle movement and $ \alpha,\beta,\gamma,\theta,$ and $\zeta $
are the components of the coordinates.

The Chapman-Enskog procedure is an asymptotic expansion method for solving
the Boltzmann equation in kinetic theory.
It is necessary to introduce a small parameter in an asymptotic expansion to
compare orders. The lattice constants are required to be small compared
with macroscopic characteristic length scale, for example, the edge length of
the cavity, $L$, i.e. $ \mid {\bf e}_{\sigma i} \mid /L \ll 1 $. Using
$\epsilon $ as a measure of this small scale, instead of using a unit lattice
constant and time step as in Section 2, the lattice Boltzmann equation is
written as
\begin{equation}
f_{\sigma i}({\bf x}+\epsilon {\bf e}_{\sigma i},t+\epsilon )
-f_{\sigma i}({\bf x},t)
= \Omega_{\sigma i},
\label{eq:a1}
\end{equation}
where $\Omega_{\sigma i}$ is the collision operator and $\epsilon $ is solely
used for distinguishing different orders.
The lattice Boltzmann BGK equation is
\begin{equation}
f_{\sigma i}({\bf x}+\epsilon {\bf e}_{\sigma i},t+\epsilon )
-f_{\sigma i}({\bf x},t)
=-\frac{1}{\tau} [f_{\sigma i}({\bf x},t)-f_{\sigma i}^{(0)}({\bf x},t)],
\label{eq:a2}
\end{equation}
where $f_{\sigma i} (\sigma=0\;\; i=1;\;\; \sigma=1,2 \;\; i=1,\cdots, 4)$
is the single-particle distribution function,
$f_{\sigma i}^{(0)}({\bf x},t) $ is the equilibrium distribution at
${\bf x}, t$ and $\tau $ is the single relaxation time.
 A general form of $f_{\sigma i}^{(0)}({\bf x},t) $ can be taken as
\begin{equation}
f_{\sigma i}^{(0)}({\bf x},t)=A_{\sigma}+B_{\sigma}({\bf e}_{\sigma i}\cdot
{\bf u})+C_{\sigma}({\bf e}_{\sigma i}\cdot {\bf u})^2+D_{\sigma} u^2.
\label{eq:a3}
\end{equation}
Here $A_{\sigma}, B_{\sigma}, C_{\sigma}$ and $D_{\sigma}$ are cofficients
that need to be found and depend on $\rho$, but not on $u$.
Equation (\ref{eq:a3}) can be thought as a special type of small velocity
(up to the $u^2$ term) expansion of $f_{\sigma i}^{(0)}$.

It is obvious that $B_0=C_0=0$ and Eq.~(\ref{eq:a3}) can be written separately
in the form
\begin{eqnarray}
f_{0 1}^{(0)}({\bf x},t)=A_{0}\mbox{\hspace{1.8 in}} + D_{0} u^2,\nonumber \\
f_{1 i}^{(0)}({\bf x},t)=A_{1}+B_{1}({\bf e}_{1 i}\cdot
{\bf u})+C_{1}({\bf e}_{1 i}\cdot {\bf u})^2+D_{1} u^2,\nonumber \\
f_{2 i}^{(0)}({\bf x},t)=A_{2}+B_{1}({\bf e}_{2 i}\cdot
{\bf u})+C_{2}({\bf e}_{2 i}\cdot {\bf u})^2+D_{2} u^2,
\label{eq:a4}
\end{eqnarray}
where  $f_{\sigma i}^{(0)}$ and $f_{\sigma i}$ satisfy the following
constraints:
\begin{equation}
\sum_{\sigma} \sum_{i} f_{\sigma i}=\sum_{\sigma} \sum_{i} f_{\sigma i}^{(0)}
 =\rho,
\label{eq:a5}
\end{equation}
and
\begin{equation}
 \sum_{\sigma} \sum_{i} f_{\sigma i}{\bf e}_{\sigma i}
=\sum_{\sigma} \sum_{i} f_{\sigma i}^{(0)} {\bf e}_{\sigma i}=\rho {\bf u}.
\label{eq:a6}
\end{equation}
These constraints mean that the non-equilibrium distributions do not contribute
to the local values of density and momentum.
Using Eqs.~(\ref{eq:a5}), (\ref{eq:a6}), (\ref{eq:l1}) and (\ref{eq:l4}), some
constraints for coefficients
$A_{\sigma},B_{\sigma},C_{\sigma}$ and $D_{\sigma}$ are found to be
\begin{equation}
A_0+4A_1+4A_2=\rho,
\label{eq:a11}
\end{equation}
\begin{equation}
2C_1+4C_2+D_0+4D_1+4D_2=0,
\label{eq:a12}
\end{equation}
and
\begin{equation}
2B_1+4B_2=\rho.
\label{eq:a13}
\end{equation}

Starting from the LBE with BGK collision operator, the Navier-Stokes equations
can be recovered. Taking a Taylor expansion of Eq.~(\ref{eq:a2})
gives
\begin{equation}
f_{\sigma i}({\bf x}+\epsilon {\bf e}_{\sigma i},t+\epsilon )
-f_{\sigma i}({\bf x},t)
=\sum_{n=0}^{\infty} \frac{\epsilon^n}{n!}[\frac{\partial}{\partial t}+
({\bf e}_{\sigma i} \cdot \nabla)]^n f_{\sigma i}({\bf x},t),
\label{eq:a15}
\end{equation}
where \[{\bf e}_{\sigma i} \cdot \nabla=e_{\sigma i \alpha} \partial_{\alpha}
+e_{\sigma i \beta} \partial_{\beta} \]
and the Einstein summation convention is used.
If terms up to $O(\epsilon^2)$ are retained in Eq~(\ref{eq:a15}), the results
is
\begin{eqnarray}
\epsilon[ \frac{\partial }{\partial t}+
({\bf e}_{\sigma i} \cdot \nabla)]f_{\sigma i}
+ \frac{\epsilon^2}{2}[\frac{\partial}{\partial t}+
({\bf e}_{\sigma i} \cdot \nabla)]^2 f_{\sigma i}+O(\epsilon^3) \nonumber \\
=-\frac{1}{\tau} [f_{\sigma i}({\bf x},t)-f_{\sigma i}^{(0)}({\bf x},t)].
\label{eq:a16}
\end{eqnarray}
Next, the Chapman-Enskog-like expansion is applied to Eq.(\ref{eq:a16}).
Expanding $f_{\sigma i}$ about $ f_{\sigma i}^{(0)}$ gives
\begin{equation}
f_{\sigma i}=\sum_{n=0}^{\infty}\epsilon^n f_{\sigma i}^{(n)}
= f_{\sigma i}^{(0)}+\epsilon  f_{\sigma i}^{(1)}+\epsilon^2 f_{\sigma i}^{(2)}
+ \cdots ,
\label{eq:a17}
\end{equation}
with constraints
\begin{equation}
\sum_{\sigma} \sum_i f_{\sigma i}^{(0)} \left[ \begin{array}{c}
                                               1\\ {\bf e}_{\sigma i}
                                              \end{array}  \right]
                                      =\left[ \begin{array}{c}
                                               \rho\\ \rho {\bf u}
                                              \end{array}  \right],
\label{eq:a18}
\end{equation}
and
\begin{equation}
\sum_{\sigma} \sum_i f_{\sigma i}^{(n)} \left[ \begin{array}{c}
                                               1\\ {\bf e}_{\sigma i}
                                              \end{array}  \right]
                                      =0,\;\;\; n\geq 1.
\label{eq:a19}
\end{equation}
The collision operator, $\Omega_{\sigma i} $, becomes
\begin{equation}
-\frac{1}{\tau}[\epsilon  f_{\sigma i}^{(1)}+\epsilon^2 f_{\sigma i}^{(2)}
+ \cdots ].
\label{eq:a20}
\end{equation}
To discuss changes in different time scales, three time scales,
$t_0,t_1,t_2$ are introduced as
\[t_0=t,\;\;\; t_1=\epsilon t,\;\;\; t_2=\epsilon^2 t, \]
where
\begin{equation}
\frac{\partial}{\partial t}
=\frac{\partial}{\partial t_0}
+ \epsilon \frac{\partial}{\partial t_1}
+ \epsilon^2 \frac{\partial}{\partial t_2}.
\label{eq:a21}
\end{equation}
Substituting Eqs.~(\ref{eq:a17}), (\ref{eq:a20}), and (\ref{eq:a21}) into
Eq~(\ref{eq:a16}), the equation to order of $\epsilon $ is:
\begin{equation}
 (\partial_{ t_0}+{\bf e}_{\sigma i} \cdot \nabla) f_{\sigma i}^{(0)}
=-\frac{1}{\tau}f_{\sigma i}^{(1)}.
\label{eq:a22}
\end{equation}
The equation to order of $\epsilon ^2 $ is:
\begin{equation}
\partial_{ t_0} f_{\sigma i}^{(1)}+\partial_{ t_1} f_{\sigma i}^{(0)}+
({\bf e}_{\sigma i} \cdot \nabla )f_{\sigma i}^{(1)}
+\frac{1}{2}[\frac{\partial}{\partial t_0}+({\bf e}_{\sigma i}
 \cdot \nabla )]^2 f_{\sigma i}^{(0)}= -\frac{1}{\tau}f_{\sigma i}^{(2)}.
\label{eq:a23}
\end{equation}
Using Eq.~(\ref{eq:a22}), one obtains
\begin{equation}
 (\partial_{ t_0}+{\bf e}_{\sigma i} \cdot \nabla)^2 f_{\sigma i}^{(0)}
=-\frac{1}{\tau}(\partial_{ t_0}+{\bf e}_{\sigma i} \cdot \nabla)
 f_{\sigma i}^{(1)}.
\label{eq:a24}
\end{equation}
Substituting Eq.~(\ref{eq:a24}) into Eq.~(\ref{eq:a23}) leads to
\begin{equation}
\partial_{ t_1}f_{\sigma i}^{(0)}+(1-\frac{1}{2 \tau})(\partial_{ t_0}
+{\bf e}_{\sigma i} \cdot \nabla) f_{\sigma i}^{(1)}
=-\frac{1}{\tau}f_{\sigma i}^{(2)}.
\label{eq:a25}
\end{equation}
 To derive the equations for $\rho$ and $\rho {\bf u}$ to the first order in
$\epsilon$, a summation of Eq.~(\ref{eq:a22}) with respect to $\sigma$ and $i$
is taken to give
\[\partial_{t_0}\sum_{\sigma} \sum_{i}f_{\sigma i}^{(0)}+\sum_{\sigma} \sum_{i}
({\bf e}_{\sigma i} \cdot \nabla)f_{\sigma i}^{(0)}
=-\frac{1}{\tau}\sum_{\sigma} \sum_{i}f_{\sigma i}^{(1)}=0,\]
which is the first order continuity equation
\begin{equation}
\partial_{t_0}\rho+\nabla \cdot (\rho {\bf u})=0.
\label{eq:a26}
\end{equation}
Similarly, multiplying ${\bf e}_{\sigma i}$ in Eq.~(\ref{eq:a22}) and taking
summation as above gives
\[\partial_{t_0}\sum_{\sigma} \sum_{i}f_{\sigma i}^{(0)}{\bf e}_{\sigma i}
+\sum_{\sigma} \sum_{i}({\bf e}_{\sigma i} \cdot \nabla)f_{\sigma i}^{(0)}
{\bf e}_{\sigma i}
=-\frac{1}{\tau}\sum_{\sigma} \sum_{i}f_{\sigma i}^{(1)}{\bf e}_{\sigma i}=0,\]
which can be simplified to
\begin{equation}
\partial_{t_0}(\rho{\bf u})+\nabla \cdot \sum_{\sigma} \sum_{i}
({\bf e}_{\sigma i}{\bf e}_{\sigma i})f_{\sigma i}^{(0)}=0.
\label{eq:a27}
\end{equation}
Defining the momentum flux tensor as
\begin{equation}
\Pi= \sum_{\sigma} \sum_{i}
({\bf e}_{\sigma i}{\bf e}_{\sigma i})f_{\sigma i},
\label{eq:a28}
\end{equation}
Eq.~(\ref{eq:a27}) can be rewritten as
\begin{equation}
\partial_{t_0}(\rho{\bf u})+\nabla \cdot \Pi^{(0)}=0.
\label{eq:a30}
\end{equation}
Similarly, the equations of order of $\epsilon^2 $ for $\rho$ and ${\bf u}$ can
be obtained from Eq.~(\ref{eq:a25}) as
\[ \partial_{t_1}\rho+(1-\frac{1}{2 \tau})\partial_{t_0}\sum_{\sigma} \sum_{i}
f_{\sigma i}^{(1)}+(1-\frac{1}{2 \tau}) \nabla \cdot
\sum_{\sigma} \sum_{i} f_{\sigma i}^{(1)}{\bf e}_{\sigma i}
=-\frac{1}{\tau}\sum_{\sigma} \sum_{i}f_{\sigma i}^{(2)}=0, \]
and
\[ \partial_{t_1}(\rho{\bf u})+(1-\frac{1}{2 \tau})\partial_{t_0}\sum_{\sigma}
\sum_{i} f_{\sigma i}^{(1)}{\bf e}_{\sigma i}+(1-\frac{1}{2 \tau}) \nabla \cdot
\sum_{\sigma} \sum_{i}{\bf e}_{\sigma i}{\bf e}_{\sigma i}f_{\sigma i}^{(1)}
=-\frac{1}{\tau}\sum_{\sigma} \sum_{i}f_{\sigma i}^{(2)}{\bf e}_{\sigma i}=0.
\]
Here the constraint given by Eq.~(\ref{eq:a18}) is used. By applying another
constraint, Eq.~(\ref{eq:a19}), these two equations can be simplified to
\begin{equation}
\partial_{t_1}\rho=0,
\label{eq:a31}
\end{equation}
and
\begin{equation}
\partial_{t_1}(\rho{\bf u})+(1-\frac{1}{2 \tau})\nabla \cdot \Pi^{(1)}=0.
\label{eq:a32}
\end{equation}
Write $\Pi^{(0)}$ as follows:
\begin{eqnarray}
 \Pi_{\alpha \beta}^{(0)}= \sum_{\sigma} \sum_{i} e_{\sigma i \alpha}
e_{\sigma i \beta}f_{\sigma i}^{(0)}\mbox{\hspace{2.8 in}} \nonumber \\
=\sum_{\sigma} \sum_{i} e_{\sigma i \alpha} e_{\sigma i \beta}
(A_{\sigma} +B_{\sigma}e_{\sigma i \gamma}u_{\gamma}+C_{\sigma}
e_{\sigma i \gamma} e_{\sigma i \theta}u_{\gamma}u_{\theta}+D_{\sigma} u^2)
\mbox{\hspace{0.5 in}} \nonumber \\
=\sum_{\sigma}A_{\sigma}2 e_{\sigma}^2\delta_{\alpha \beta}
+\sum_{\sigma}C_{\sigma}u_{\gamma}u_{\theta} \sum_{i}e_{\sigma i \alpha}
e_{\sigma i \beta}e_{\sigma i \gamma}e_{\sigma i \theta}
+\sum_{\sigma}D_{\sigma}2 e_{\sigma}^2\delta_{\alpha \beta} u^2 \nonumber \\
=(2A_1+4A_2)\delta_{\alpha \beta}+C_1 u_{\gamma}u_{\theta}
(2 \delta_{\alpha \beta \gamma \theta})
\mbox{\hspace{2 in}} \nonumber \\
+C_2 u_{\gamma}u_{\theta}(4 \Delta_{\alpha \beta \gamma \theta}
-8 \delta_{\alpha \beta \gamma \theta})
+(2D_1 u^2+4D_2 u^2)\delta_{\alpha \beta}.\mbox{\hspace{1 in}}\nonumber
\end{eqnarray}
Since
\[u_{\gamma}u_{\theta} \delta_{\alpha \beta \gamma \theta}
=u_{\alpha}u_{\beta}\delta_{\alpha \beta}, \]
and
\[ 4u_{\gamma}u_{\theta} \Delta_{\alpha \beta \gamma \theta}
=4(\delta_{\alpha \beta}\delta_{\gamma \theta}
+\delta_{\alpha \gamma} \delta_{\beta \theta}
+\delta_{\alpha \theta} \delta_{\beta \gamma})u_{\gamma}u_{\theta} \\
=4\delta_{\alpha \beta} u^2+8u_{\alpha}u_{\beta}, \]
one can write $\Pi_{\alpha \beta}^{(0)}$ as
\begin{equation}
\Pi_{\alpha \beta}^{(0)}=[2A_1+4A_2+(4C_2+2D_1+4D_2)u^2]\delta_{\alpha \beta}
\nonumber \\
+8C_2u_{\alpha}u_{\beta}+(2C_1-8C_2)u_{\alpha}u_{\beta}\delta_{\alpha \beta}.
\label{eq:a33}
\end{equation}
The first term is the pressure term and the other two are nonlinear terms.
In order to obtain velocity-independent pressure, the coefficient of $u^2$ is
chosen to satisfy
\begin{equation}
4C_2+2D_1+4D_2=0.
\label{eq:a34}
\end{equation}
To have Galilean invariance, the non-isotropic term is eliminated by choosing
\begin{equation}
2C_1-8C_2=0.
\label{eq:a35}
\end{equation}
Eq.~(\ref{eq:a33}) becomes
\begin{equation}
\Pi_{\alpha \beta}^{(0)}=(2A_1+4A_2)\delta_{\alpha \beta}
+8C_2u_{\alpha}u_{\beta}.
\label{eq:a36}
\end{equation}
Assuming that
\begin{equation}
8C_2=\rho,
\label{eq:a37}
\end{equation}
and
\begin{equation}
2A_1+4A_2=c_s^2 \rho,
\label{eq:a38}
\end{equation}
where $c_s$ is speed of sound, gives the final expression for $\Pi^{(0)}$ as
\begin{equation}
\Pi_{\alpha \beta}^{(0)}=c_s^2 \rho \delta_{\alpha \beta}
+\rho u_{\alpha}u_{\beta}.
\label{eq:a39}
\end{equation}
Substituting Eq.~(\ref{eq:a39}) into Eq.~(\ref{eq:a30}) results in
\begin{equation}
\frac{\partial (\rho {\bf u})}{\partial t_0}+\nabla \cdot (\rho {\bf u}{\bf u})
=-\nabla (c_s^2 \rho).
\label{eq:a40}
\end{equation}
Eqs.~(\ref{eq:a26}) and (\ref{eq:a40}) are Euler equations that are derived
from the $\epsilon$-order of the expansion of the Boltzmann equation.

To derive the equations accurate to $\epsilon ^2$, the quantity $\nabla \cdot
\Pi^{(1)}$ needs to be evaluated. From Eq.~(\ref{eq:a22}) the non-equilibrium
distribution can be expressed as
\begin{equation}
f_{\sigma i}^{(1)}=-\tau \partial_{t_0} f_{\sigma i}^{(0)}-\tau (e_{\sigma i
\gamma} \partial_{\gamma})  f_{\sigma i}^{(0)}.
\label{eq:a41}
\end{equation}
Substituting $f_{\sigma i}^{(1)}$ into $\Pi_{\alpha \beta}^{(1)}$ gives
\begin{eqnarray}
\Pi_{\alpha \beta}^{(1)}=\sum_{\sigma} \sum_{i} e_{\sigma i \alpha}
e_{\sigma i \beta}f_{\sigma i}^{(1)}
=-\tau \partial_{t_0}\sum_{\sigma} \sum_{i} e_{\sigma i \alpha}
e_{\sigma i \beta}f_{\sigma i}^{(0)}-\tau \partial_{\gamma}
\sum_{\sigma} \sum_{i} e_{\sigma i \alpha}
e_{\sigma i \beta}e_{\sigma i \gamma}f_{\sigma i}^{(0)} \nonumber \\
=-\tau \{\partial_{t_0}\Pi_{\alpha \beta}^{(0)}+\partial_{\gamma}
\sum_{\sigma} \sum_{i} e_{\sigma i \alpha}e_{\sigma i \beta}e_{\sigma i \gamma}
(A_{\sigma}+B_{\sigma}e_{\sigma i \theta}u_{\theta}+C_{\sigma}
e_{\sigma i \theta}e_{\sigma i s}u_{\theta}u_s+D_{\sigma} u^2) \}.\nonumber
\end{eqnarray}
Using Eq.~(\ref{eq:a39}) for $\Pi_{\alpha \beta}^{(0)}$ leads to
\begin{eqnarray}
\Pi_{\alpha \beta}^{(1)}=-\tau \{\partial_{t_0}[(c_s^2 \rho)
\delta_{\alpha \beta}+\rho u_{\alpha} u_{\beta}]
+ \partial_{\gamma}B_1 u_{\theta}2 \delta_{\alpha \beta \gamma \theta}
+ \partial_{\gamma}B_2 u_{\theta}(4 \Delta_{\alpha \beta \gamma \theta}
-8\delta_{\alpha \beta \gamma \theta}) \} \nonumber \\
=-\tau \{-c_s^2\delta_{\alpha \beta} \partial_{\gamma}(\rho u_{\gamma})
+\partial_{t_0}(\rho u_{\alpha} u_{\beta})+\partial_{\alpha}
(2B_1-8B_2)u_{\beta}\delta_{\alpha \beta} \mbox{\hspace{1 in}} \nonumber \\
+4\partial_{\gamma}(B_2 u_{\gamma})\delta_{\alpha \beta}
+4\partial_{\alpha}(B_2 u_{\beta})+4\partial_{\beta}(B_2 u_{\alpha})\}.
\mbox{\hspace{2 in}}
\label{eq:a42}
\end{eqnarray}
To avoid non-isotropy, set
\begin{equation}
2B_1-8B_2=0.
\label{eq:a43}
\end{equation}
Recalling Eq.~(\ref{eq:a13}), $B_1$ and $B_2$ can be uniquely determined as
\begin{equation}
B_2=\frac{\rho}{12},\;\;\; B_1=\frac{\rho}{3}.
\label{eq:a44}
\end{equation}
Therefore, Eq.~(\ref{eq:a42}) can be written as
\begin{equation}
\Pi_{\alpha \beta}^{(1)}=-\tau \{\frac{1}{3}\partial_{\gamma}
(\rho u_{\gamma})\delta_{\alpha \beta}+\frac{1}{3}\partial_{\alpha}
(\rho u_{\beta})
 +\frac{1}{3}\partial_{\beta}(\rho u_{\alpha})
-c_s^2\partial_{\gamma}(\rho u_{\gamma})\delta_{\alpha \beta}
+\partial_{t_0}(\rho u_{\alpha} u_{\beta})\}.
\label{eq:a45}
\end{equation}
The last term can be simplified using Eq.~(\ref{eq:a40}) to take the form
\begin{eqnarray}
\partial_{t_0}(\rho u_{\alpha} u_{\beta})= u_{\beta}\partial_{t_0}
(\rho u_{\alpha})+\rho u_{\alpha}\partial_{t_0} u_{\beta}\mbox{\hspace{3.1 in}}
\nonumber \\
= u_{\beta}[-\partial_{\gamma}(\rho u_{\alpha}u_{\gamma})-\partial_{\alpha}
(c_s^2\rho)]+ u_{\alpha}\partial_{t_0}(\rho u_{\beta})- u_{\alpha}u_{\beta}
\partial_{t_0}\rho \mbox{\hspace{2 in}}\nonumber \\
=- u_{\beta}\partial_{\gamma}(\rho u_{\alpha}u_{\gamma})- u_{\beta}
\partial_{\alpha}(c_s^2\rho)
- u_{\alpha}[\partial_{\gamma}(\rho u_{\beta} u_{\gamma})+\partial_{\beta}
(c_s^2\rho)]+ u_{\alpha} u_{\beta}\partial_{\gamma}(\rho u_{\gamma})
\mbox{\hspace{0.8 in}}\nonumber \\
=- u_{\alpha}\partial_{\beta}(c_s^2\rho)- u_{\beta}\partial_{\alpha}
(c_s^2\rho)- u_{\alpha}u_{\beta}\partial_{\gamma}(\rho u_{\gamma})
-\rho  u_{\alpha}u_{\gamma}\partial_{\gamma}u_{\beta}
+ u_{\alpha}u_{\beta} \partial_{\gamma}(\rho u_{\gamma})
-u_{\beta}\partial_{\gamma}(\rho u_{\alpha} u_{\gamma}) \nonumber \\
=-u_{\alpha}\partial_{\beta}(c_s^2\rho)-u_{\beta}\partial_{\alpha}
(c_s^2\rho)-\rho  u_{\alpha}u_{\gamma}\partial_{\gamma}u_{\beta}
-u_{\beta}\partial_{\gamma}(\rho u_{\alpha} u_{\gamma}).\mbox{\hspace{1.8 in}}
\nonumber
\end{eqnarray}
Eq.~(\ref{eq:a45}) therefore becomes
\begin{eqnarray}
\Pi_{\alpha \beta}^{(1)}=-\tau \{\frac{1}{3}\partial_{\gamma}
(\rho u_{\gamma})\delta_{\alpha \beta}+\frac{1}{3}\partial_{\alpha}
(\rho u_{\beta}) \nonumber +\frac{1}{3}\partial_{\beta}(\rho u_{\alpha})
-c_s^2\partial_{\gamma}(\rho u_{\gamma})\delta_{\alpha \beta} \nonumber \\
-u_{\alpha}\partial_{\beta}(c_s^2\rho)-u_{\beta}\partial_{\alpha}
(c_s^2\rho)-\rho  u_{\alpha}u_{\gamma}\partial_{\gamma}u_{\beta}
-u_{\beta}\partial_{\gamma}(\rho u_{\alpha} u_{\gamma})\}.
\label{eq:a46}
\end{eqnarray}
Note that $\Pi_{\alpha \beta}^{(1)}$ does not only depend on the first
spatial derivatives of $\rho$ and ${\bf u}$.
Ignoring the last two terms which are order of $O(u^3)$ in Eq.~(\ref{eq:a46}),
leads to
\begin{eqnarray}
\Pi_{\alpha \beta}^{(1)}=-\tau \{(\frac{1}{3}-c_s^2)\partial_{\gamma}
(\rho u_{\gamma})\delta_{\alpha \beta}+\frac{1}{3}\partial_{\alpha}
(\rho u_{\beta}) \nonumber \\
+\frac{1}{3}\partial_{\beta}(\rho u_{\alpha})
-u_{\alpha}\partial_{\beta}(c_s^2\rho)-u_{\beta}\partial_{\alpha}
(c_s^2\rho) \} +O(u^3).
\label{eq:a48}
\end{eqnarray}
Combine equations of $O(\epsilon)$ and $O(\epsilon^2)$ for $\rho$ and
${\bf u}$, and Eqs.~(\ref{eq:a26}), (\ref{eq:a40}), (\ref{eq:a31})
and  (\ref{eq:a32}) with  Eq.~(\ref{eq:a48}) as follows:\\
 Eq.~(\ref{eq:a26}) added to Eq.~(\ref{eq:a31}) multipled by $\epsilon $ gives
\[\partial_{t_0}\rho + \epsilon \partial_{t_1} \rho + \nabla \cdot
(\rho {\bf u})=0, \]
which gives the correct form of the continuity equation as
\begin{equation}
\frac{\partial \rho}{\partial t} + \nabla \cdot (\rho {\bf u})=0.
\label{eq:a49}
\end{equation}
 Eq.~(\ref{eq:a40}) added to Eq.~(\ref{eq:a32}) multipled by $\epsilon $ gives
\begin{equation}
\partial_{t}(\rho {\bf u})+ \nabla \cdot (\rho {\bf u}{\bf u})=
- \nabla \cdot (c_s^2 \rho)-\epsilon (1-\frac{1}{2 \tau}) \nabla \cdot
\Pi_{\alpha \beta}^{(1)}.
\label{eq:a50}
\end{equation}
Substituting Eq.~(\ref{eq:a48}) for $\Pi_{\beta \alpha}^{(1)}$,
Eq.~(\ref{eq:a50}) becomes
\begin{eqnarray}
\partial_{t}(\rho u_{\alpha})+\partial_{\beta}(\rho u_{\alpha} u_{\beta})
=- \partial_{\alpha} (c_s^2 \rho)+\epsilon (\tau-\frac{1}{2})
\partial_{\beta} \{(\frac{1}{3}-c_s^2) \partial_{\gamma}(\rho u_{\gamma})
\delta_{\alpha \beta} \\
+ \frac{1}{3}\partial_{\alpha}(\rho u_{\beta})
+\frac{1}{3}\partial_{\beta}(\rho u_{\alpha})-u_{\alpha}\partial_{\beta}
(c_s^2 \rho)-u_{\beta}\partial_{\alpha}(c_s^2 \rho) \}
+O(u^3)+O(\epsilon^3),
\end{eqnarray}
which may be written in the form
\begin{eqnarray}
\partial_{t}(\rho u_{\alpha})+\partial_{\beta}(\rho u_{\alpha} u_{\beta})
=- \partial_{\alpha}(c_s^2 \rho)+\epsilon (\tau-\frac{1}{2}) \{
\partial_{\alpha}[(\frac{1}{3}-c_s^2) \partial_{\gamma}(\rho u_{\gamma})]
\nonumber \\
+\partial_{\beta}[\frac{1}{3}\rho(\partial_{\alpha} u_{\beta}+
\partial_{\beta}u_{\alpha})]+\partial_{\beta}[(\frac{1}{3}-c_s^2)
(u_{\alpha}\partial_{\beta}\rho+u_{\beta}\partial_{\alpha}\rho)] \}
+O(u^3)+O(\epsilon^3).
\label{eq:a52}
\end{eqnarray}
Consider the constraints on $A_{\sigma}$ given in Eqs.~(\ref{eq:a11}) and
(\ref{eq:a38}), and choosing
\[ A_0=\frac{4}{9} \rho, \;\;\; A_1=\frac{1}{9}\rho, \;\;\;
A_2=\frac{1}{36}\rho,\]
Eq.~(\ref{eq:a11}) is satisfied and the sound speed is determined by
\[ c_s^2=\frac{1}{3}. \]
 Eq.~(\ref{eq:a52}) is simplified as
\begin{equation}
\partial_{t}(\rho u_{\alpha})+\partial_{\beta}(\rho u_{\alpha} u_{\beta})
=- \partial_{\alpha}(c_s^2 \rho)+\frac{1}{3}\epsilon (\tau-\frac{1}{2})
\partial_{\beta}[\rho (\partial_{\alpha}u_{\beta}+\partial_{\beta}(u_{\alpha}
)] +O(u^3)+O(\epsilon^3).
\label{eq:a53}
\end{equation}
Define the stain-rate tensor as
\begin{equation}
S_{\alpha \beta}=\frac{1}{2}(\partial_{\alpha}u_{\beta}+\partial_{\beta}
u_{\alpha}).
\label{eq:a54}
\end{equation}
Then Eq.~(\ref{eq:a53}) can be rewritten as follows:
\begin{equation}
\partial_{t}(\rho u_{\alpha})+\partial_{\beta}(\rho u_{\alpha} u_{\beta})
=- \partial_{\alpha}(c_s^2 \rho)+2\nu \partial_{\beta}
(\rho S_{\alpha \beta})+O(u^3)+O(\epsilon^3),
\label{eq:a55}
\end{equation}
where
\begin{equation}
\nu=\frac{2 \tau -1}{6} \epsilon ,
\label{eq:a56}
\end{equation}
with $\nu $ being the  the kinematic viscosity.
Recall the Navier-Stokes equations in the two-dimensional space [18]
\begin{equation}
\partial_{t}(\rho u_{\alpha})+\partial_{\beta}(\rho u_{\alpha} u_{\beta})
=- \partial_{\alpha}p+ \partial_{\beta} \{ 2\mu (S_{\alpha \beta} - \frac{1}
{2} u_{\gamma \gamma}\delta_{\alpha \beta}) \},
\label{eq:a57}
\end{equation}
and
\begin{equation}
\partial_{t}\rho +\nabla \cdot (\rho {\bf u})=0.
\label{eq:a58}
\end{equation}
For an incompressible fluid with constant viscosity, $\mu$, the Navier-Stokes
equations become
\begin{equation}
\partial_{t}(\rho u_{\alpha})+\partial_{\beta}(\rho u_{\alpha} u_{\beta})
=- \partial_{\alpha}p+ \partial_{\beta} \{ 2\mu S_{\alpha \beta} \}
=- \partial_{\alpha}p + \mu  \partial_{\beta} \partial_{\beta}u_{\alpha}.
\label{eq:a59}
\end{equation}
For $\rho=$ constant, the incompressible Navier-Stokes equations are
\begin{eqnarray}
\nabla \cdot {\bf u}=0,\mbox{\hspace{1.7 in}} \nonumber \\
\partial_{t} u_{\alpha}+\partial_{\beta}(u_{\alpha} u_{\beta})
=- \partial_{\alpha}(\frac{p}{\rho})+ \nu \partial_{\beta}^2
u_{\alpha }.
\label{eq:a60}
\end{eqnarray}
It is seen that Eq.(\ref{eq:a55}) is exactly the same as the Narier-Stokes
equation (\ref{eq:a57}) in the incompressible limit, $\nabla \cdot {\bf u}=0$.

Collecting all coefficients so far, one obtains
\[A_0=\frac{4}{9} \rho, \;\; A_1=\frac{1}{9} \rho,\;\; A_2=\frac{1}{36} \rho,\]
\[B_1=\frac{1}{3} \rho,\;\; B_2=\frac{1}{12} \rho, \]
\[C_1=\frac{1}{2} \rho,\;\; C_2=\frac{1}{8} \rho. \]
The remaining coefficients $D_0,D_1$ and $D_2$ are related by
Eq.~(\ref{eq:a12}) and Eq.~(\ref{eq:a34}), so there is one free parameter.
Since all coefficients of particle 2 are one-fourth of the corresponding
coefficients of particle 1, one can require $D_1=4 D_2$. Hence,the remaining
coefficients are determined as
\[D_0=-\frac{2}{3}\rho,\;\; D_1=-\frac{1}{6}\rho,\;\; D_2=-\frac{1}{24}\rho,\]
Finially, the equilibrium distribution functions are given as
\begin{eqnarray}
f_{0 1}^{(0)}=\frac{4}{9}\rho[1-\frac{3}{2}u^2], \mbox{\hspace{1.65 in}}
\nonumber\\
f_{1 i}^{(0)}=\frac{1}{9}\rho[1+3({\bf e}_{1 i}\cdot
{\bf u})+\frac{9}{2}({\bf e}_{1 i}\cdot {\bf u})^2-\frac{3}{2}u^2],\nonumber \\
f_{2 i}^{(0)}=\frac{1}{36}\rho[1+3({\bf e}_{2 i}\cdot
{\bf u})+\frac{9}{2}({\bf e}_{2 i}\cdot {\bf u})^2-\frac{3}{2}u^2].
\label{eq:a61}
\end{eqnarray}

\end{document}